\documentclass[12pt]{article}
\usepackage{latexsym}
\usepackage{graphics}
\usepackage{epsfig}
\usepackage[mathscr]{eucal}
\usepackage{amscd}
\usepackage{amssymb}
\usepackage{amsmath}
\usepackage{amssymb}

\newcommand{\qed}{\hfill $\Box$}
\newcommand{\re}{\mathbb R}
\newcommand{\bZ}{Z}

\newcommand{\bN}{\mathbb N}

\newcommand{\nin}{\notin}
 
\newcommand{\vl}{\mathsf v}
\newcommand{\tvl}{\tilde \vl}

\newcommand{\cM}{\mathsf M}
\newcommand{\cF}{\mathcal F}

\newcommand{\cO}{\mathcal {O}}

\newcommand{\cI}{\mathcal I}

\newcommand{\fut}{\mathrm{Fut}}
\newcommand{\past}{\mathrm{Past}}

\newcommand{\alex}{\mathbf I}

\newcommand{\vol}{\mathrm{vol}}

\newcommand{\cA}{\mathcal A} 
\newcommand{\can}{A}

\newcommand{\cP}{\mathcal P}
\newcommand{\hv}{\mathsf n}

\newcommand{\cT}{{{\mathsf T}}}
\newcommand{\ccT}{\mathcal T}
 
\newcommand{\be}{\begin{equation}}
\newcommand{\jj}{\mathbf J}

\newcommand{\ee}{\end{equation}}

\newcommand{\pprec}{\prec\!\prec}

\newcommand{\cN}{\mathcal N}

\newtheorem{theorem}{Theorem}
\newtheorem{lemma}{Lemma}
\newcommand{\ssuc}{\succ\!\succ} 
\newcommand{\wb}{\pprec_{\mathsf b}} 
\newcommand{\wa}{\ssuc_{\mathsf a}}

\newcommand{\cU}{\mathcal U} 
 
\newcommand{\cx}{\mathfrak r}
 
\newcommand{\tcN}{\tilde \cN} 

\title{Causal Set Topology}

\author{Sumati Surya \\
Raman Research Institute, Bangalore, India } 
\begin{document}

\maketitle
\begin{abstract}

The Causal Set Theory (CST) approach to quantum gravity is motivated
by the observation that, associated with any causal spacetime $(M,g)$
is a poset $(M,\prec)$, with the order relation $\prec$ corresponding
to the spacetime causal relation. Spacetime in CST is assumed to have
a fundamental atomicity or discreteness, and is replaced by a locally
finite poset, the causal set. In order to obtain a well defined
continuum approximation, the causal set must possess the { requisite}
intrinsic topological and geometric properties that characterise a
continuum spacetime in the large.  The continuum approximation thus 
sets the stage for the study of topology in CST.  
We review the status of causal set topology and present some new
results relating poset and spacetime topologies.  The hope is that in
the process, some of the ideas and questions arising from CST will be
made accessible to the larger community of computer scientists and
mathematicians working on posets.

\end{abstract} 

\section{The Spacetime Poset and Causal Sets} 
In Einstein's theory of gravity, space and time are merged into a
single entity represented by a pseudo-Riemannian or Lorentzian
geometry $g$ of signature $-+++$ on a differentiable four dimensional
manifold $M$. Unlike a Riemannian space in which the distance between
two distinct points is always positive definite, a Lorentzian geometry
is characterised by the fact that this distance can be positive
definite (spacelike), negative definite (timelike) or zero (null).

The {\sl classical}\footnote{ The term classical is used to indicate
that theory does not involve quantum effects.}  theory of gravity is
determined by the Einstein field equations,  a non-linear set
of second order partial differential equations which relate the
spacetime geometry $g$ to classical non-gravitational or matter fields
\footnote{The electromagnetic field is an important example of a
  classical matter field.}.  Geometries which obey the Einstein field
  equations include simple ones like flat, or Minkowski spacetime, as
  well as exotic ones containing blackholes.  To date, the theory is
  observationally sound, being consistent with observations spanning
  lengths scales from the orbits of geocentric satellites to the
  expansion of the observable universe. At smaller length scales,
  including human scales and all the way down to nuclear scales, the
  effects of spacetime curvature are deemed negligible, which means
  that Einstein gravity can be functionally replaced by Newtonian
  gravity.

Despite this satisfactory performance however, the theory is not
without serious theoretical flaws. Spacetimes which satisfy Einstein's
equations can contain ``singular'' points at which the geometry is
ill-defined or the associated curvature diverges, an example being the
initial ``big-bang'' singularity of an expanding universe. The structure
of a classical spacetime no longer sustains near such a point and
hence the classical geometry is inadequate to answer the question --
what is the nature of the singularity?  Moreover, at atomic and
nuclear scales, the matter fields themselves become quantum in
character, and one has to replace Einstein's equations with an
ill-understood hybrid in which classical spacetime interacts with
quantum matter. Results from the hybrid theory challenge several
cherished notions in physics, a prime example being the so-called
``information loss paradox'' in the presence of blackholes
\cite{Hawking}.  They suggest the existence of an underlying, fully
quantum theory of gravity in which spacetime itself is ``quantised''.

What the nature of such a fundamental theory should be is anybody's
guess, since there are almost no observational constraints on the
theory except those coming from its limits to both non-gravitational
quantum theory ($ G \rightarrow 0$) and Einstein gravity ($\hbar
\rightarrow 0$).  The different approaches to quantum gravity adopt
different cocktails of physical principles and include string theory,
loop quantum gravity, spin foams, dynamical triangulations, causal set
theory and others.  However, a universal characteristic of any such
theory which mixes quantum (characterised by the Planck constant
$\hbar = 10^{-34} J.s $) with gravity (characterised by the
gravitational constant $G = 6.67 \times 10^{-11} m^3/kg/s^2$ and the
speed of light in vacuum $c= 3 \times 10^{8} m/s $), is the existence
of a fundamental length scale, the Planck scale $\ell_p = \sqrt{G\hbar
c^{-3}} \sim 10^{-35} m$, 20 orders of magnitude smaller than the
nuclear scale. 

The interest of the present work lies in the poset based approach of
causal set theory (CST) \cite{bometal}. The physical principles that
guide this approach are causality, local Lorentz invariance and the
assumption of a fundamental spacetime atomicity. CST posits that the
underlying structure of spacetime is a locally finite partially
ordered set, the {\sl causal set} with the order relation
corresponding to the so-called causal relation in the continuum
approximation.

\begin{figure}[ht]
\centering \resizebox{2.0in}{!}{\includegraphics{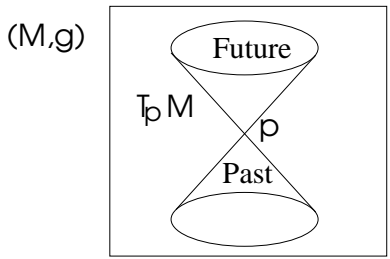}}
\vspace{0.5cm}
\caption{{\small The past and future lightcones in the tangent space
    $T_pM$ at a point $p$ in a spacetime $(M,g)$.}}
\label{lightcones.fig}
\end{figure}

To motivate this approach, let us look more closely at the properties
of a Lorentzian spacetime $(M,g)$.  Because of the signature $-+++$,
the set of future timelike directions at any $p \in M$ forms a {\sl
future light cone} in the tangent space $T_pM$, with $p$ at its apex,
and similarly the set of past timelike directions, a {\sl past light
cone}, as in Fig \ref{lightcones.fig}.
The boundary of these lightcones are the future and past null
directions. For a $C^1$ curve $\gamma$ on $M$, if the tangent vector
$\xi(p)$ at every $p \in \gamma$ is always either spacelike, timelike
or null, then $\gamma$ is defined to be a timelike, spacelike or null
curve, respectively. $\gamma$ is said to be {\sl causal} if its
tangent is everywhere either timelike or null. Causal curves represent
pathways for physical signals, i.e., those that travel at or slower
than the speed of light.  Causal curves on $M$ give rise to a causal
order $\prec$ on $M$: For $x,y \in M$, $x\prec y $ if there is a
future directed causal curve from $x$ to $y$. Moreover, in any
spacetime, every point has a neighbourhood $U \subset M$ such that
$(U,\prec_U)$ is a partial order, where $\prec_U$ is determined by
causal curves that lie in $U$.

While the causal order is always transitive, it need not be
acyclic. Spacetimes can contain so-called ``closed causal curves'',
favoured by science fiction writers to build time-machines. Indeed,
Einstein theory admits such acausal solutions, an example being the
G\"{o}del universe \cite{HE}. Spacetimes for which the causal order is
acyclic are said to be {\sl causal}, which makes $(M,\prec)$ a partial
order.  For strongly causal spacetimes, it was shown by Malament
\cite{malament} that the poset $(M,\prec)$ determines the conformal
class of $(M,g)$ with the only remaining freedom given by the local
volume element\footnote{Two geometries represented by the metrics
$g_1,g_2$ are said to belong to the same conformal class if
$g_1=\Omega^2g_2$, where $\Omega^2$ is a nowhere vanishing function on
$M$. In $n$ spacetime dimensions, the metric $g$ is an $n \times n$
symmetric matrix and hence the causal structure determines all but one
of its $n(n+1)/2$ independent components.}. This can be summarised by
the statement ``Causal structure+ Volume = Spacetime''.

The CST approach to quantisation gives prime importance to this poset
structure of a causal spacetime in the spirit of the Malament result.
The assumption of a fundamental spacetime atomicity or
discreteness\footnote{In this paper, unless explicitly stated,
discrete will mean atomistic or non-continuum and not the
trivial/discrete topology.}  in CST means that not all of $(M,\prec)$
is considered relevant to the underlying quantum theory but only a
subset thereof. Such a subset should have the property that regions of
finite spacetime volume contain a finite number of elements of the
underlying causal set.  Discreteness is posited as a cure for the
divergences of Einstein's theory and quantum field theories, and is
supported by the existence of a fundamental Planck volume $V_p$
suggesting that structures substantially smaller than $V_p$ should not
be relevant to the theory. The continuum emerges in the large as an
approximation, in a manner similar to a continuous stream of water
approximating the underlying discreteness of water molecules. This
analogy is useful for picturing the continuum approximation, which
will be defined more rigourously later in this section.  Discreteness
is encoded in CST by the assumption of local finiteness, i.e., that
the intervals $<x,y> = \{z \in P| x \preceq z \preceq y\} $ have
finite cardinality. $(M,\prec)$ is therefore not itself a causal set
but contains as a subposet a causal set which approximates to $(M,g)$.
Thus,  a CST version of Malament's result reads ``Order +
Cardinality $\approx$  Spacetime''.  This correlation
between the spacetime volume and causal set cardinality in the
continuum approximation is a key feature of CST.

As an aside, we note that the idea of causality and local finiteness
are also natural to a poset model of computation. Computations can be
modelled within a non-relativistic framework, with signals assumed to
travel at infinite speed, and with an absolute notion of simultaneity.
Causality then reduces to a simple arrow of time. However, since
signal velocity is physically constrained, there is indeed a ``signal
velocity cone'' similar to the light cone in a spacetime which, roughly
speaking, separates events which can influence each other and those
that cannot. Any associated distance function will then have a
Lorentzian character. Local finiteness, on the other hand, comes from
the physical constraint on the number of computations per interval of
time. Thus, in a very general sense, a model of computation based on
causality and local finiteness is, at the least, a kinematical
realisation of a causal set.

\begin{figure}[ht]
\centering \resizebox{2.0in}{!}{\includegraphics{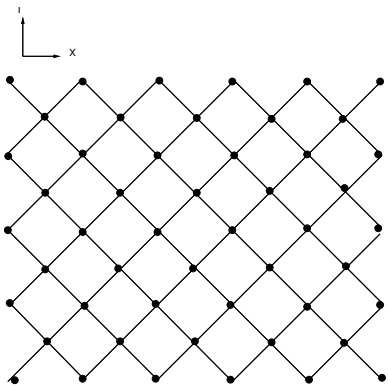}}
\vspace{0.5cm}
\caption{{\small Diamond lattice in a preferred coordinate system}}
\label{lattice.fig}
\end{figure}

As noted above, $(M,\prec)$ is not itself a causal set, which means
that $(M,g)$ needs to be discretised to obtain the underlying causal
set $C$, keeping in mind the number to volume correspondence.  The
simplest possible discretisation of a Riemannian space is one that is
{ regular}, with a fixed number of lattice points per unit volume. An
example of this is the square lattice on the 2 dimensional Euclidean
plane.  However, regularity is a deceptive concept in the
discretisation of a Lorentzian spacetime. Take for example
2-dimensional Minkowski spacetime with the metric $ds^2 = -c^2dt^2 +
dx^2$, where $(t,x)$ are the time and space coordinates, and $c$ is
the speed of light. A diamond lattice with equally spaced lattice
points along constant $u=ct-x$ and $v=ct+x$ null directions appears
regular in the $(t,x)$ coordinate system, with an apparently fixed
volume to number correspondence, as in Fig \ref{lattice.fig}. Here,
the order relation on the lattice is the induced causal order.
\begin{figure}[ht]
\centering \resizebox{2.0in}{!}{\includegraphics{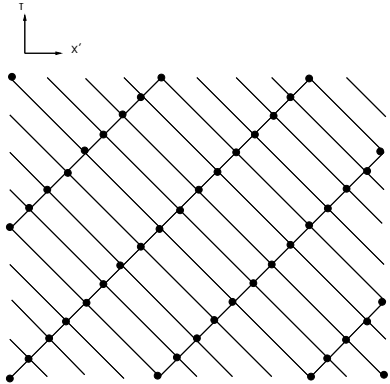}}
\vspace{0.5cm}
\caption{{\small  Diamond lattice under a boost.}}
\label{boosted.fig}
\end{figure}
However, the lattice is no longer regular when acted upon by a
``boost'' transformation of the coordinates, $B:(t,x) \rightarrow
(t',x')$, the Lorentzian analog of a rotation in Euclidean
space. Here, $t'=\gamma(t-vx/c^2)$, $x'=\gamma(x-vt)$, where
$\gamma={\sqrt{1-v^2/c^2}}^{-1}$ and $v < c$ is the relative speed
between the two coordinate systems. In Euclidean space, a rotation
about the origin takes a point $(x_1,y_1) $ to another point $
(x_2,y_2)$ on the constant $r=\sqrt{x_1^2+y_1^2}= \sqrt{x_2^2+y_2^2}$
circle centred at the origin. Similarly, a boost takes $(t_1,x_1) $ to
another point $ (t_2,x_2)$ on the constant
$\rho^2=c^2t_1^2-x_1^2=c^2t_2^2-x_2^2$ hyperbola. Thus, under a boost,
the diamond lattice transforms to one in which the lattice points are
squeezed together in one null direction and pulled apart in an
orthogonal null direction as shown in Fig \ref{boosted.fig}. A large
boost with $|v/c| $ close to 1 gives rise to large spacetime ``voids'',
which violate the number to volume correspondence.  In contrast, large
voids cannot result from a rotation of a regular lattice in Euclidean
space.

How important is invariance under a boost or, {\sl Lorentz invariance}
(LI)\footnote{Lorentz invariance is a symmetry only of Minkowski
spacetime. However, a local version of it, referred to as local
Lorentz invariance, is a property of all spacetimes. We will be sloppy
and use the acronym LI to refer to this local version whenever
necessary.}?  Several approaches to discretisation allow for Lorentz
violation \cite{SF}. However, LI is a cherished fundamental symmetry
of nature, and one which has been experimentally tested to a high
degree of accuracy \cite{LI}. In order to incorporate it into CST
without forfeiting the volume-number correspondence, one has to define
the discretisation more carefully.  Again, the Euclidean analogy is
instructive. For a Euclidean lattice, the continuous rotational
symmetry group is replaced by a discrete subgroup which picks out
preferred directions. Instead, for a random lattice obtained, for
example, by randomly sprinkling points on the plane, there are no
preferred directions.  Similarly, in CST, spacetime is discretised by
a Poisson sprinkling of points in $(M,g)$ which then ensures that
there are no preferred directions and hence no Lorentz violation
\cite{bomhensor}.  For a Poisson process the probability of sprinkling
$n$ points in a spacetime region of volume $V$ is $P_V(n)=
\frac{1}{n!}  \exp^{-V/V_c} (V/V_c)^n $, $V_c$ being the
discretisation scale. Here $<n>=V/V_c$,  which means that the volume to
number correspondence is satisfied, though only in the mean.

Thus, in order to maintain LI and a reasonable semblance of the
volume-number correspondence, the continuum approximation of CST must
incorporate a random process. We now define this CST continuum
approximation. A causal set $C$ is said to be {\sl approximated} by a
spacetime $(M,g)$ if there is an embedding map $\Phi:(C,\prec)
\rightarrow (M,g)$ which is {\sl faithful}\footnote{We follow
Bombelli's definition for a faithful embedding \cite{luca}.}: Let $C$
be a finite causal set with cardinality $ V/V_c$ and $M$ a finite
volume $V$ region of a spacetime.  Let $V_0$ be the volume of sampling
intervals in $M$, with $V_c< V_0 < V$. If $I$ denotes a spacetime
interval of volume $V_0$, the indicator function $F_n= \int \chi_n(I)
dI/\int dI$ is such that $\chi_n(I)=1$, or $0$ depending on whether
$I$ contains $n$ elements of the embedded causal set $\Phi(C)$ or not,
with the integral over all possible $I$'s in $(M,g)$. Then, if $|F_n -
P_{V_0}| < \delta$, $\Phi$ will be said to be a {\sl $\delta$-faithful
embedding} with respect to $V_0$. Faithfully embeddable will
henceforth be used in this $(\delta, V_0)$ sense.  In order to make
the definition compatible with our intuition, we will also require
that $V_c \ll V_0 \ll V$ and $0< \delta \ll 1 $.  For suitable choices
of $\delta$ and $V_0$, a causal set generated by a Poisson sprinkling
into $M$ with density $V_c^{-1}$ is faithfully embedded in $M$.  A
regular spacetime lattice on the other hand, will not faithfully embed
for any reasonable choice of $(\delta, V_0)$.

In CST, one considers the set of all causal sets, not only those that
are obtained via a discretisation of a spacetime. However, in order to
make a connection with Einstein theory, causal sets that approximate
to spacetimes play a special role. If a causal set is indeed the
appropriate discretisation of a continuum spacetime and can capture
its large scale structure, then large scale spacetime geometry and
topology must be encoded in purely order theoretic terms in the
causet. Moreover, if the classical limit of CST is Einstein theory, 
causal set geometry and topology must play a non-trivial role in CST
dynamics. This forms the main motivation for the present work.

If CST is to yield meaningful results, it is also important that the
continuum approximation be unique, at least at scales larger than the
discreteness scale. This is the key conjecture of causal set theory,
also known as its ``Hauptvermutung''. It states that if
$\Phi:(C,\prec) \rightarrow (M,g)$ is a faithful embedding at density
$V_c^{-1}$, then $(M,g)$ is unique upto isomorphisms at scales above
$V_c$.  Results in the literature support the idea behind the
conjecture, \cite{bommeyer,meyer, bg} and include recent work on the
closeness of Lorentzian spaces \cite{bomnoldus}. A rigorous statement
and proof however, would require a better understanding of how causal
set topology and geometry relate to that of the approximating
spacetime.

We begin the next section with some definitions from Lorentzian
geometry and posets, which will serve as a rough dictionary between
the two. In \cite{MP} it was shown that a globally hyperbolic
spacetime is a bicontinuous poset whose interval topology is the
manifold topology. We present an alternative proof, based on a
different set of assumptions. This suggests that the condition of
global hyperbolicity may be replaced by a weaker causality condition.
We then show that, for a causal set that approximates to a causal
spacetime $(M,g)$, the chronological relation and the way-below
relation coincide. In the following section, we review Sorkin's
arguments for a ``finitary'' topology \cite{fintop}.  Using a theorem
due to McCord \cite{mccord} we show that for a suitable choice of
cover, Sorkin's finitary structure suffices to capture the continuum
topology upto weak homotopy equivalence
\footnote{The author would like to thank Jimmy Lawson for help with 
reference \cite{mccord}.}.  In the following section we review 
results on a finitary construction in CST based on inextendible 
antichains \cite{hompaper}. This construction reproduces the continuum
homology for causal sets that approximate to globally hyperbolic 
spacetimes and gives important supporting evidence for a topological
version of the CST Hauptvermutung.  We conclude with some of the open
questions on causal set topology. 

\section{Topology from Causal Structure}

By a {\sl poset} $P$ we mean a set with an order relation $\prec$
which is (i) {\sl transitive}, i.e., $x \prec y$ and $ y \prec z \Rightarrow
x \prec z$ and (ii) {\sl acyclic}, i.e., $ x \prec y, y \prec x \Rightarrow
x =y $, for any $x,y,z \in P$. Acyclicity is also implied by the
{\sl irreflexive} condition $x \nprec x$, a convention used in much of the
causal set literature. We will avoid it here, since it makes the map
between the causal set and the continuum a little more cumbersome. 
The {\sl past} and {\sl future} sets of an element $x \in P$ are defined as
$\past(x)\!=\! \{y| y\prec x \}$, $\fut(x)\!=\!\{y| x \prec y
\}$. Without the irreflexive condition therefore, $x$ belongs to its
own past and future.  A {\sl causal set} is a poset which is also
locally finite: (iii) $|\past(x) \cap \fut(y)| < \infty$, where $|A|$
denotes the cardinality of the set $A$.

While all differentiable manifolds, being paracompact, admit metrics
of positive definite signature, not all admit Lorentzian metrics. We
will only concern ourselves here with those that do.  As described in
the introduction, in a Lorentzian spacetime $(M,g)$, the tangent space
$T_pM$ at $p \in M$ is divided into timelike, null and spacelike
vectors, which, because of the $-+++$ signature form lightcones. We
will henceforth assume here that all spacetimes are time-orientable,  so
that a consistent notion of future and past directed vector fields is
possible everywhere on  $(M,g)$. 

The causal past and future of $x $ are defined as $\jj^-(x) = \{ y | y
\prec x \} $, $\jj^+(x) = \{ y | x \prec y \}$, respectively.  Note
that since a causal vector can have zero norm, $x \prec x$. The poset
analog of an interval in $(M,\prec)$ is then the set $\jj^+(x) \cap
\jj^-(y)$. 

The Lorentzian signature means that spacetimes are endowed with
another order relation: $x \pprec y$ if there exists a $C^1$ curve
from $x $ to $y$ whose tangent is everywhere future timelike. Again,
$\pprec$ is transitive, and for a causal spacetime, also
acyclic. However, $\pprec $ is irreflexive since $x$ cannot link to
itself via a curve whose tangent is everywhere strictly of negative
norm. The associated {\sl chronological} past and future of $x$ are
defined as $\alex^-(x)= \{ y | y \pprec x \}$ and $\alex^+(x)= \{ y |
x \pprec y \}$, respectively. In the manifold topology, $\alex^\pm(x)$
are open sets, while $\jj^\pm(x)$ is closed only for a limited class
of spacetimes. A {\sl generalised transitivity} condition states that
$x \pprec y \prec z \Rightarrow x \pprec z$, and $x \prec y \pprec z
\Rightarrow x \pprec z$.

As pointed out in \cite{MP}, the chronological relation $\pprec$ has
an analog in posets in the { way below} relation $\wb$ and its dual
the { way above} relation $\wa$.  This relation is defined in terms of
directed sets, which appear in domain theory \cite{domain}. In a  {\sl
directed set} $S$  for every pair $s,s' \in S \, \, $,
there is an $ s'' \in S$ such that $s,s' \prec s''$. A filtered set
is defined dually. Simple spacetime examples of a directed set are
$\alex^-(x)$, which does not contain its supremum $x$, and $\jj^-(x)$
which does. However, directed sets are more general than such past
sets, and can contain disconnected pieces, an example being a pair of
past directed non-intersecting time-like curves from $x$.  For $x,y
\in P$, $x$ is said to be {\sl way below} $y$ or $x \wb y$, if for all
directed sets $S$ with a supremum $\bigsqcup S$ and $y \prec \bigsqcup
S$, $\exists \, \, s \in S$ such that $ x \prec s $. The $\wb$
relation is a subsidiary relation to $\prec$, similar to how the
chronological relation is subsidiary to the causal relation.

The {\sl Alexandroff interval} in a spacetime is defined to be the
open set $\alex(x,y) \equiv \alex^+(x) \cap \alex^-(y)$.  The
Alexandroff intervals form a basis for the {\sl Alexandroff topology},
and it is a well established result in Lorentzian geometry that in a
strongly causal spacetime\footnote{Strong causality refers to a causal
spacetime in which causal curves cannot reenter an arbitrarily small
neighbourhood of a point.}, the Alexandroff topology is the same as
the manifold topology \cite{Penrose}.  The appropriate poset analog to
an Alexandroff interval would then appear to be the interval based on
the $\wb$ relation, $\ll x, y\gg = \{ z \in P| x \wb z \wb y\}$. For
bicontinuous posets\footnote {A {\sl continuous} poset is one in which
the set of all elements way below $x$ contains an increasing sequence
with supremum $x$ and implies that arbitrary interpolations are
admissible, i.e., for all $x \prec y $ in $P$ $\exists \, \, z \in P $
such that $x \prec z \prec y$. {\sl Bicontinuity} refers to a
continuous poset for which the order reverse conditions are true, and
such that $x \wb y \Rightarrow y \wa x$.}, these intervals form the
basis for the so-called {\sl interval topology} on the poset.

Indeed, in \cite{MP} it was shown that for globally hyperbolic
spacetimes\footnote{Spacetimes in which $\overline{\alex(x,y)}$ is
compact for all $x, y \in M$ are called globally hyperbolic, and
represent classically well behaved geometries. In the ``hierarchy'' of
conditions on the causal structure, strong causality is one of the
weakest, while global hyperbolicity is one of the most stringent
conditions \cite{Penrose,HE}.}  $\pprec$ and $\wb$ do coincide
in $(M,\prec)$ and hence so do the intervals $\ll x,y \gg$ and
$\alex(x,y)$.  Moreover, $(M, \prec)$ {\it is} a bicontinuous poset,
which means that the interval topology on $(M, \prec)$ coincides with
the manifold topology of $(M,g)$.  The results of \cite{MP} are
striking, in that the {\it full} manifold topology of a globally
hyperbolic spacetime $(M,g)$ is captured in purely order theoretic
terms by the poset $(M,\prec)$. This supplements the result of
Malament\cite{malament} and places the order theoretic motivations for
CST on a firmer footing.

It is therefore of interest to know if the results of \cite{MP} hold
in spacetimes satisfying weaker causality conditions than global
hyperbolicity. The following results  are an attempt in this
direction. While Lemma \ref{lemmaone.lem} holds for any causal
spacetime, Lemma \ref{lemmatwo.lem} replaces the global hyperbolicity
condition with a topological requirement on directed sets.

\begin{lemma} 
For a causal spacetime, the $\wb$ relation in $(M,\prec)$ implies the
chronological relation in $(M,g)$. \label{lemmaone.lem} 
\end{lemma} 
 
\noindent {\bf Proof:} Let $x \wb y$. If we take the directed set
$S=\alex^-(y)$, then $\sqcup S = y$, and $\, \, \exists \, \, s \in
\alex^-(y)$ such that $x \preceq s$. But any such $s \pprec y$ so that
$x \pprec y$ \cite{Penrose}. \qed.

This generalises a part of the proof of \cite{MP} to all causal
spacetimes, not only those that are globally hyperbolic. To prove the
converse, a compactness condition on $\overline{\alex(x,y)}$ was used
in \cite{MP} which requires the spacetime to be globally hyperbolic
and not just causal. Instead, if we impose on $(M,\prec)$ the
topological condition that for all directed sets $S$, every
neighbourhood of $\sqcup S$ has non-trivial intersection with $S$,
then we can show the following:

\begin{lemma} 
In a causal spacetime $(M,g)$, if for all directed sets $S$ in
$(M,\prec)$ every neighbourhood of $\sqcup S$ intersects $S$
non-trivially, then the chronological relation in $(M,g)$ implies the
$\wb$ relation in $(M,\prec)$. \label{lemmatwo.lem} 
\end{lemma} 

\noindent {\bf Proof:} Let $x \pprec y$. Let $S$ be a directed set
with $ y\prec \sqcup S $. Then, by generalised transitivity, $x \pprec
\sqcup S$. Since $\alex^+(x)$ is open, this means that there exists a
neighbourhood $N$ of $\sqcup S$ such that $N \subset \alex^+(x)$. From
the assumptions on directed sets in $(M,\prec)$, $N\cap S \neq
\emptyset$ and hence there exists an $s \in S$ such that $x \pprec
s$. Since this is true for all directed sets, $x \wb y$. \qed

Thus, for any spacetime which is causal and satisfies this topological
restriction on directed sets specified in Lemma \ref{lemmatwo.lem},
the way below relation is the chronological relation. As argued in
\cite{MP}, if one further imposes the condition of strong causality,
the interval topology on $(M,\prec)$ agrees with the manifold topology
on $(M,g)$.  Whether this condition on the topology of directed sets
translates into a causality condition weaker than global hyperbolicity
is currently under investigation \cite{PS}.

The spacetime $(M,g)$ admits other topologies defined via the order
relations $\prec$ and $\pprec$, distinct from the manifold topology.
For past complete spacetimes, consider the open sets to be $\alex^+(x)
$. It is easily seen that the associated topology is $T_0$: for any $x
\prec y$ and every $w\in M $ such that $x \in \alex^+(w)$, $ y \in
\alex^+(w)$, so that every open set containing $x$ also contains
points in its chronological future. However, for every $x \prec y $, 
$\,\, \exists \,\, z \in M$ such that $y \in \alex^+(z) $, but $x \nin
\alex^+(z)$.  Finally, for $x,y$ spacelike to each other, there exist
a pair $w \pprec x $, $z \pprec y$ such that $x \in \alex^+(w), y \nin
\alex^+(w)$ and $y \in \alex^+(z), x \nin \alex^+(z)$. This topology
is therefore not $T_1$ and hence insufficient to describe the
Hausdorff topology on $M$\footnote{ Assuming global hyperbolicity, the results
of \cite{MP} moreover tell us that $\{ \alex^+(x)\} $ form a basis for
the so-called {\sl Scott topology}, since $\alex^+(x)$ are upper sets
and because they are open in the manifold topology, are ``inaccessible
by directed suprema''}.

We have so far steered clear of the condition that makes a poset a
causal set, namely, local finiteness. We remind the reader that the
poset of interest to CST is {not} $(M, \prec)$ itself, but the causal
set $(C,\prec)$ which faithfully embeds into $(M,g)$. $C$ is therefore
not continuous, and hence the results of \cite{MP} are not immediately
applicable to CST. 
However, local finiteness simplifies some things.  In particular, if
$\Phi: (C, \prec) \rightarrow (M,g)$ is a faithful embedding, $\wb$ in
$C$ is indeed equivalent to the chronological relation in $\Phi(C)
\subset M$, as we now show.

First, we note that for any directed set $S$ with supremum $\sqcup S$
in a locally finite poset $C$, $\sqcup S \in S$ and is therefore a
maximal element. The proof is simple. Assume that $\sqcup S \nin
S$. Let $s \in S$, so that the interval $<s,\sqcup S)> \neq \emptyset
$ and let its cardinality be $a < \infty$ (by local finiteness). Let
$s_1 \in S\cap \fut(s) $, so that $<s_1, \sqcup S> $ has cardinality
$a_1 < a$. Next, let $s_2 \in S \cap \fut(s_1)$ such that $<s_2,
\sqcup S>$ has cardinality $a_2 < a_1$ and so on. This gives a finite,
exhaustive set $\{s_1, s_2 \ldots s_k\}$ so that $a_k=2$, i.e., $<s_k,
\sqcup S>$ contains only $s_k$ and $\sqcup S$.  Now, for any $s' \in
S$, there exists an $s'' \in S $ such that $s''\succ s_k$ and $s''
\succ s'$. This means that $s''=s_k$, or that $s_k$ is to the future
of all $s' \in S$ and hence the supremum of $S$, which is a
contradiction.

\begin{lemma} 
For a causal set $C$ that faithfully embeds into a causal spacetime
$(M,g)$, the relation $\wb $ in $C$ is the same as the chronological
relation in $\Phi(C) \subset M$.
\end{lemma}

\noindent {\bf Proof:} Let $x \wb y$ in $C$.  Choose $S=\past(y)$, so
that $\sqcup S = y \in S$.  Then there exists an $s \in S$ such that
$x \prec s$. Since $ s \prec y $ this means that $x \prec y$. For a
causal set that embeds into a spacetime it is therefore possible that
$y$ lies on the future lightcone of $x$, or that they can only be
joined by a future directed null curve.  However, for a faithful
embedding, given an $x \in M$, the probability for this is zero
\footnote{ This is for the same reason that the probability for any
$y$ to lie in a given lower dimensional submanifold in $M$ is zero in 
a Poisson sprinkling.}.  Thus, with probability one, $x \pprec y$.  To
prove the converse is easier than in the continuum, because of the
simplicity of directed sets in $C$.  If $ x \pprec y$, and $S$ a
directed set with $y \prec \sqcup S$, then since $\sqcup S \in S$, $x
\pprec y \prec \sqcup S$, $\Rightarrow x \pprec \sqcup S$ and hence $x
\prec \sqcup S$, so that $x \wb y$. \footnote{These arguments also
  show that for a locally finite poset, the way below relation is
  equivalent to the causal relation.} \qed

Thus, the intervals $\ll x,y\gg$ in $C$ correspond to Alexandroff
intervals $\alex(x,y)$ in $(M,g)$. However, since not all Alexandroff
intervals in $(M,g)$ are generated by $\ll x,y\gg$ in $C$, the
interval topology does not have a simple relationship to the manifold
topology.  Indeed, in the continuum approximation, only ``relevant''
continuum topology can be obtained from the causal set.  One way to do
this might be to consider a subcover of the Alexandroff topology which
obtains from $C$, and ask how much of the topology of $(M,g)$ it
encodes. This question was posed in a more general context by Sorkin
in \cite{fintop}, by replacing the continuum by a finitary
substitute. This is the topic of the following section.

\section{Finitary Topology} 

The idea of the continuum is based on the assumption that an infinite
resolution of the events in a spacetime is possible, at least in
principle. Even without considering quantum effects, this assumption
begs the question: given the limited resolution of measurement devices
how much of a continuum spacetime (assuming it exists) do we actually
probe, and what is its relation to the continuum?  In \cite{fintop} it
was suggested that the answer to this question may lie in replacing
the continuum topology with a subtopology of locally
finite(LF)\footnote{We shall use this acronym to avoid confusion with
local finiteness of posets.}  open coverings. This yields a
``finitary'' substitute for the full topology, which we can use to
address the above question.

We now review this construction.  Let $\cO = \{ O_i \} $ be an LF 
cover of $X$, i.e., every $x \in X$ has a neighbourhood $N \ni
x$ such that $N \cap O_i \neq \emptyset$ for only a finite number of
$O_i$.  This space can be made into a $T_0$ space by defining the
equivalence $x \sim y$ if $ x \in O_i \Leftrightarrow y \in O_i$ for
all $O_i \in \cO $. The ``finitary'' quotient space $\cF = X/ \sim $
carries the induced topology $\ccT_\cF$ on $\cF$ with the sets $\{ f_
1, f_2 \ldots f_k\} \subset \cF$ being open if their inverses under
the quotient map are open in $\cO$. For every $f_1,f_2 \in \cF, f_1
\neq f_2$, there exists an open $U \in \ccT_\cF$, such that either $f_1
\in U, f_2 \nin U$ or vice versa. Thus, $\cF$ can be endowed with a
$T_0$ topology.  The worry of all discretisation procedures is how
much relevant continuum information is lost in the process. For
example, would $\ccT_\cF$ have sufficient information to be able to
recreate the topological invariants of $X$?

In \cite{fintop} it was shown that this $T_0$ quotient does contain
non-trivial topological information, by defining, further, a partial
order $P(\cF)$ from $\ccT_\cF$. The neighbourhood of any
$[x] \in \cF$ is defined as $\Lambda([x])= \cap \{U \in \ccT_\cF| [x]
\in U \}$. Then, $[x] \prec [y]$ in $P(\cF)$ iff $ \Lambda([x])
\subset \Lambda([y])$. This poset can be endowed with non-trivial
topologies. As shown in \cite{fintop}, in certain examples, the chain
complex on $P(\cF)$ is homotopic to $X$, thus suggesting a potential
connection between finitary topology and that of $X$. The { full}
topology of a bounded $T_1$ space $X$ is recovered by taking a
directed collection of finite open covers $\cO_i$ of $X$, so that the
finitary spaces $\cF(\cO_i)$ converge to $X$ as the $\cO_i$ become
more refined. In this sense, $X$ is recovered in the limit of infinite
refinement.

However, if discreteness is thought to be fundamental, the recovery of
what we will call ``irrelevant'' topology in the limit is not what we
are after. Rather, we would wish for the discrete structure to retain
all the relevant features of the continuum without having to go to the
limit of infinite refinement. Consider for example a space $X$ with
topological structure at scales far smaller than the available
resolution. These structures are irrelevant to coarser observations.
From the point of view of a measurement, $X$ is equivalent to a space
$Y$ which has no topological structures at scales of the order of the
discreteness scale.

As we will now show, an appropriate initial choice of $\cO$ is indeed
sufficient and one does not need to invoke the limit of infinite
refinement.  Let $P(\cO) $ be a poset obtained from the relation of
set inclusion for a point finite open covering $\cO$ of $X$. Let
$K(\cO)$ be the associated chain simplicial complex and $|K(\cO)|$ the
underlying polyhedron with the weak topology.  We invoke the following
theorem due to McCord \cite{mccord}: 
\begin{theorem}({\bf McCord}) If $\cO$ is a point finite basis-like
  open cover of $X$ by homotopically trivial sets, then there exists a
  weak homotopy equivalence \footnote{ $f:
  X \rightarrow Y$ is said to be weak homotopy equivalent if the
  induced maps $f^i_\ast: \pi_i(X,x) \rightarrow \pi_i(Y,f(x))$ are
  isomorphisms for all $x \in X$ and all $i \geq 0$, where $\pi_i$
  denotes the $ith$ homotopy group.} $f: |K(\cO)| \rightarrow X$.
\end{theorem}

Although the poset $P(\cO)$ differs in general from that obtained from
the finitary space $P(\cF)$, we show that they are indeed equivalent
for coverings which are ``sparse'' enough. This allows us to use
McCord's theorem to establish a weak homotopy equivalence between
$P(\cF)$ and $X$.

\begin{lemma} 
Let $\cO$ be a point finite basis-like simple cover of $X$, such that
none of the  $O_i \in \cO$ can be expressed as a non-trivial union of
members of $\cO$. Then, there exists a map $g: P(\cF) \rightarrow
P(\cO)$ which is an order preserving bijection, with the order
relation in $P(\cF)$ carrying over to the order relation on
$P(\cO)$. Hence, the chain complex $K(\cF)$ of $P(\cF)$ is
such that there exists a weak homotopy equivalence $f: |K(\cF)|
\rightarrow X$. \label{lem}
\end{lemma} 

\noindent {\bf Proof:} Let $g([x])= \cap_{\cO} \{O \in \cO| x \in
  O\}\equiv O(x) $, i.e. $[x]$ is mapped to the smallest open set in
  $\cO$ containing $x$. For any $y \in X $, if $O(y) \neq O(x)
  \Rightarrow x \nsim y$. Therefore $g$ is 1-1 and we may write
  $O([x])\equiv O(x)$.  Conversely, associated with each $O$, there is
  a finite sequence of equivalence classes $\{[x_1],[x_2], \ldots
  [x_k]\}$ such that every $x \in O$ lies in some equivalence class
  $[x_i]$. Let $\{O(x_1), \ldots, O(x_k)\}$ be the associated open
  sets. Then there exists an $x_i \in O$ such that $O(x_i) = O$. Else,
  $O=\cup_i O(x_i)$ and does not belong to the collection $\cO$. In
  other words, for every $O \in \cO$ is associated a unique $[x]$ in
  $\cF$. Thus, $g$ is a bijection. That it is further, order
  preserving, comes from the fact that if $[x] \prec [y]$ in $P(\cF)$,
  $\Lambda([x]) \subset \Lambda([y])$, while in $P(\cO)$, $O(x) \prec
  O(y)$ iff $ O(x) \subset O(y)$. Now, under the quotient map,
  $f(O(x))= \Lambda([x])$. This can be seen as follows. $[x] \in U
  (\in \cT_\cF) \Leftrightarrow x \in f^{-1}(U)$, $O(x) \subseteq
  f^{-1}(\Lambda([x])$ since otherwise there exists a smaller
  $O'=f^{-1}(U) \cap O(x) \subset O(x)$ which contains $x$. Assume
  strict inclusion, i.e., $O(x) \subset f^{-1}(\Lambda([x])$. Then
  $\exists $ a set $\{[z_1], [z_2] \ldots [z_k]\} \in \cF$ such that
  $z_i \in f^{-1}(\Lambda[x]) \Rightarrow z_i \nin O(x)$. Let this 
  set be exhaustive, i.e., there is no other equivalence class $[z]$
  such that $z \in f^{-1}(\Lambda[x]), z \nin O(x)$. Thus, if
  $f(O(x))= \{[x], [y_1], \ldots [y_l] \} $ where $x,y_i \in O(x)$,
  then $\Lambda[x]= \{[x], [y_1], \ldots [y_l], [z_1] \ldots [z_k]
  \}$. Since $f(O(x)) \subset \Lambda(x)$, it contradicts the claim
  that $\Lambda(x)$ is the smallest open subset of $\ccT_\cF$
  containing $[x]$. Since the order relation in both posets is defined
  purely in terms of set inclusion, and every open set in $\cO$ is
  also a neighbourhood of some $x$, this means that $g$ is an order
  preserving bijection. It then follows from the theorem of McCord
  that $f: |K(\cO)| \rightarrow X$ is a weak homotopy
  equivalence. \qed

Thus, a carefully chosen finitary substructure suffices to capture the
relevant topological information of a space.  Assuming that the space
is topologically trivial at scales below some fixed scale, a LF
covering of open sets at this scale can capture the coarse continuum
topology.

The considerations above are independent of the signature of the
overlying geometry. For a Lorentzian geometry $(M,g)$, a natural
choice of topology would be that based on the Alexandroff intervals
out of which a sparse, simple subcover may be obtained, so that the
associated $|K(\cF)|$ is weakly homotopic to $M$. Note, however, that
the order relation for the poset $P(\cF)$ does {\it not} correspond to
the causal order. If at all a relationship exists between a causal set
and this finitary poset, it should be more subtle.

\section{Causal Set Homology} 

As shown in the previous section, a weak homotopy correspondence
exists for a given class of cover and the underlying space. However
what needs to be understood is whether CST hands us the right sort of
cover in the continuum approximation, which satisfies the required
property of LF and sparseness.  For the interval topology on $C$, the
associated Alexandroff cover on $(M,g)$ is { not} LF and hence doesn't
lend itself naturally to the finitary analysis of the previous
section. The challenge is to isolate appropriate subsets of $C$ which
provide a basis for a topology on $C$, which can then be correlated
with the continuum topology in the approximation.

In \cite{hompaper} such a candidate class of subsets was proposed and
shown to capture the continuum homology when $(M,g)$ is globally
hyperbolic.  For such spacetimes $M \sim \Sigma \times \re$, where
$\Sigma$ represents a spatial hypersurface, so that $\Sigma$ is a
deformation retract of $M$. This allows us to shift the focus away
from $M$ to $\Sigma$.  The causal set analog of $\Sigma$ is an {\sl
inextendible antichain} $A$ which is a maximal set of unordered points
in the causal set. On its own, $A$ is only endowed with the trivial
topology, and hence does not suffice to capture the topology of
$\Sigma$. Consider instead a neighbourhood of $A$ in $C$, the {\sl
thickened antichain}
\begin{equation} 
\label{thickening} 
\cT_{\hv}(A) \equiv \{ \, p \, | \, \, | (  \past(p)) 
\cap \fut(\can)| \leq {\hv}\}, \quad {\hv} \in \bN. 
\end{equation} 
Here, $\cT_1(A) = A $ and as $\hv$ increases, one gets to sample more
and more of the future of $A$ in $C$.  The past $P_i$ of a maximal
element $ m_i \in \cT_n(A)$ casts a ``shadow'' $A_i \equiv P_i \cap A
$, on $A$. For $\hv$ finite, the collection $\cP \equiv \{ P_i\} $
covers $\cT_\hv(A) $, where $i \in [1, \ldots, m]$, $m= |\cM|$, for
$\cM$ the set of maximal elements of $\cT_\hv(A)$. Similarly, the
collection $\cA \equiv \{ A_i\}$, $i \in [1, \ldots, m] $ covers
$A$. 

The covers $\cP$ and $\cA$ which occur naturally in the causal set are
thus candidates for recovering the large scale topology of the
spacetime. In particular, $\cA$ is a discrete version of an open cover
on $\Sigma$. In order to find the continuum analogue of the thickened
antichain, $\Sigma$ must be volume ``thickened'' to a region
$M_\Sigma^\vl$. The volume function is $\vl(p) \equiv
\vol(\alex(\Sigma,p)), \, \, p \in \alex^+(\Sigma)$, where $\vol(X)$
is the spacetime volume of a region $X \subset M$. The level sets
$\Sigma_\vl$ of $\vl$ are homeomorphic to $\Sigma$ and provide a
foliation of $M$ to the future of $\Sigma$. $M_\Sigma^\vl$ is thus the
region sandwiched between $\Sigma$ and $\Sigma_\vl$.  For any $\vl>0$,
the intersection of $\alex^-(x)$ with $\Sigma$ for $ x \in \Sigma_\vl$
is open in $\Sigma$. These shadows provide a cover $\cO$ of $\Sigma$,
from which a LF subcover $\cO_\vl$ can be obtained. Unlike
the discrete case, however, an LF  subcover $\cI_\vl$ of the
past sets $I(x) = \alex(\Sigma, x), x \in \Sigma_\vl$ does not cover
$M_\Sigma^\vl$.

One could at this stage construct a finitary topology of $\cO_\vl$ and
find the associated simplicial complex. Before we do this, let us
first try to guess the discrete-continuum correspondence that we are
seeking. While $\cO$ covers $\Sigma$ and are analogs of the cover
$\cA$ of $A$, these are sets of measure zero in the Poisson sprinkling
and hence one cannot directly obtain the correspondence between the
two. Instead, one must compare the collections of past sets $\cP$ and
$\cI$. In \cite{hompaper} it was shown that these collections are in
1-1 correspondence with their spatial counterparts $\cA$ and $\cO$,
respectively. An LF  collection $\cO_f$ of $\cO$ can be
obtained via the continuum pasts of the maximal elements of the
thickened antichain. In order to use the results of Lemma [\ref{lem}],
$\cO_f$ must be a point finite, sparse and simple cover. The condition
of sparseness is not natural to such a cover, however, and hence an
alternative construction is required.

In \cite{hompaper} the finitary topology used was the nerve simplicial
complex associated with a cover. In the causal set, the nerves
$\cN_\hv(\cP)$ and $\cN_\hv(\cA)$ were shown to be homotopic to each
other and so too their continuum counterparts $\cN(\cI_f)$ and
$\cN(\cO_f)$.  For a generic causal set, $\cN_\hv(\cA)$ provides a
``transient'' topology on $A$ since it is in general non-homotopic to
$\cN_{\hv'}(\cA)$ for $\hv \neq \hv'$. While the nerve $\cN(\cO_\vl)$
also varies with $\hv$, an important feature of the continuum is that
there is a large enough range of $\vl$ for which $\cN(\cO_\vl)$ is
homotopically stable.

Since $(\Sigma,h)$ is itself a Riemannian space, with $h$ the spatial
metric on $\Sigma$ induced from $g$, every $x \in \Sigma$
has a {\sl convexity radius}. This is the largest $\cx_x$ such that
the distance function is convex on the open ball $B(x,\cx_x)$ and any
two points in $B(x,\cx_x)$ are joined by unique segments lying
entirely in it \cite{petersen}. The infimum of $\cx_p$ over $p \in
\Sigma$ gives the convexity radius $\cx$ of $\Sigma$. An open set $O$
is said to be {\sl convex with respect to $\cx$} if for any $p,q \in
O$, the (unique) geodesic between them of arc-length $< \cx$ lies
entirely in $O$. A {\sl convex cover} of $\Sigma$ is an LF 
cover of open sets which are convex with respect to $\cx$.  The
following result due to de Rham and Weil \cite{dederham} then states
\begin{theorem} 
\rm{{\bf {(De Rham-Weil)}}}  \label{dederham} 
The nerve of a convex cover of $\Sigma$ is weakly homotopic to
$\Sigma$. 
\end{theorem}  
Using the machinery of causal analysis, it was shown in
\cite{hompaper} that associated to every $\Sigma$ is a {\sl convexity
volume} $\tvl$ such that for all $0 < \vl < \tvl$, the shadows from
$\Sigma_\vl$ are convex. Thus for any LF subcover $\cO_f
\subset \cO$, $\cN_\vl(\cO_f)$ is weakly homotopic to $\Sigma$.

The tricky part is to correlate this result to the nerve
$\cN_\hv(\cP)$ or $\cN_\hv(\cA)$ obtained from the causal set. Because
of the nature of the continuum approximation, the correspondence is a
probabilistic one. There are, moreover, crucial differences between
the continuum and the causal set. Namely, apart from the upper bound
$\tvl$ coming from convexity considerations, one also has a lower
bound $\vl_0$ coming from the discreteness scale. At such scales, the
transient causal set topology $\cN_\hv(\cA)$ need bear no semblance to
the continuum. In general, for small $\hv$ it will be made up of
several disconnected bits. Thus, it is only in an intermediate range
that we can expect a correspondence that will be useful in the sense
of Theorem [\ref{dederham}]. In such a range, and under certain
conditions on the compactness scale of the spacetime, $\cN_\hv (\cA)$
was shown to be, with high probability, an adequate subcomplex of a
$\cN(\cU)$ where $\cU$ are open covers induced by the chronological
pasts of the maximal elements of $\cT_\hv(A)$.

Thus if $\Phi: C \rightarrow (M,g)$ is a faithful embedding, then for
any inextendible antichain $A$ for which the discreteness scale is
sufficiently smaller than the supremum of the convexity volumes of the
set of $\Sigma \supset A$, there is a range of $\hv$ for which the
homology is stable and equivalent to the homology of
$M$. While the analytical work uses very stringent conditions to
obtain high probabilities, numerical simulations \cite{numpaper} with
relatively small causal sets already produce striking results. They
suggest a criterion for manifoldlikeness based on the existence of a
stable homology, sandwiched between fluctuating homology.

It is useful at this point to compare this construction with that of
persistent homology \cite{persistent}. For calculating the persistent
homology of a complex $K$, use is made of a filtration of subcomplexes
$\{K_i \}$ of $K$, $i=0, \ldots n$, ordered by inclusion $K_{i-1}
\subset K_i$, with $K_n =K$. In the filtration, each simplex is
preceded by its faces, so that $K$ can be expressed as an ordered set
of simplices. One has the induced homomorphism $f_p^i: H_p(K_{i})
\rightarrow H_p(K_{i+1})$, which then combines to $f_p^{ij}:
H_p(K_{i}) \rightarrow H_p(K_{j})$, $0 \leq i \leq j \leq n$.
Persistent homology then refers to the homology which survives the
filtration from the $i$th stage to the $j$th stage, $i \leq j$.
Computationally, the filtration provides an advantage, since the
homology can be incrementally updated.  The idea of persistence thus
has a resonance with the nerve construction and the stability of the
homology.

However, the set of nerve complexes $ \{ \cN(\cA_\hv) \}$, $1 \leq \hv
\leq \hv_0$ do not provide a filtration for $\cN(\cA_{\hv_0})$. This
is because for each $\hv$, the nerve is constructed from shadows of
maximal elements, so that the shadows from stage $\hv-1$ are not
necessarily contained as vertices in $\cN(\cA_\hv)$ complex.  One
could rectify this by simply adding all shadows upto stage $n$, so
that shadows from all elements in $\cT_n(A)$ are included in the new
nerve complex $\tcN(\cA_\hv)$ including the elements of $A$ itself.
This enlarges the complex significantly but provides the required
filtration. While the simplices are not added stage by stage, one
could nevertheless extract the persistent homology groups from this
construction.  However, in a sense, persistence is already intrinsic
to the thickened antichain construction, with the ordering
provided by the volume or cardinality function.  For example, if at
stage $\hv_1$ the zeroth homology $H_0= \bZ^p , p>1$ and at stage
$\hv_2$ $H_0= \bZ$, then the ``spurious'' extra components in
$\cN(\cA_{\hv_1})$ are discarded in $\cN(\cA_{\hv_2})$. Using only the
pasts of the maximal elements of $\cT_{\hv_2}(A)$ is thus sufficient
to obtain the required stable or persistent homology. While
$\cN(\cA_\hv)$ has fewer vertices than its filtered counterpart
$\tcN(\cA_\hv)$ there may be yet unexplored computational advantages
in using the persistent homology algorithms. What we can say
definitively, however, is that the vertices of a filtration of
$\cN_n(A)$ will not all have faithful continuum counterparts, since
for small thickenings, the associated spacetime regions would be of
order the discreteness scale.

\section{Open Questions} 

While there has been progress on the question of causal set topology
in the context of the CST continuum approximation, there is much that
remains to be understood. Perhaps the strongest result on topology
that we might hope for is a topological version of the Hauptvermutung:
if $\Phi_1: C \rightarrow (M_1,g_1)$ and $\Phi_2: C \rightarrow
(M_2,g_2)$ are faithful embeddings, then the topological
invariants\footnote{These include the algebraic invariants like
homotopy and homology.} of the associated finitary topologies $\cF_1,
\cF_2$ should be equal, where the $\cF_i$ on $M_i$ obtain from the
embeddings $\Phi_i$, $i=1,2$.  This would mean that the coarse topology of
$C$ uniquely determines the topology of $M$ at scales larger than the
discreteness scales.

We are clearly not close to a proof of such a conjecture, but it is
not all that far from our sights. Consider the homology construction
of the previous section.  For a $(\delta, V_0)$ pair of faithful
embeddings $\Phi_1, \Phi_2$, the $H_i$ should be constructed from
coverings using a $\vl \sim V_0$, so that the associated homologies
are equivalent. Even if one restricts to spacetimes which are globally
hyperbolic, our current homology construction does not, at least at
present, say anything about the {\it existence and ubiquity} of the
special class of inextendible antichains used therein. If it were
possible to identify or characterise these antichains with no
reference to the approximating manifold, we would be close to a
homological version of the Hauptvermutung.

In order to make such an identification, one may have to invoke
geometric structures on the causal set. Simulations suggest that the
desired antichains can be identified by the homogeneity of the
extrinsic curvature of the spatial slices which contain them.  Indeed,
the thickening procedure seems to ``flow'' any antichain to one that
is more homogeneous in this sense, thus suggesting
genericity. Progress has been made in this direction using a spatial
distance on the antichain induced by the thickening \cite{geom}.

Of course, the question of how spacetime geometry is encoded in a
causal set is itself key to CST.  In causal sets that faithfully embed
into Minkowski spacetime, it was shown in \cite{bg} that the length of
the longest antichain between two elements is a good measure of their
time-like geodesic distance. Generalising this to arbitrary
spacetimes, and finding a bound on the fluctuations would be of
considerable interest to the CST community. In order to build a causal
set dynamics that mimics that of spacetime, a causal set analog of the
curvature is required. While there are hints in this direction
\cite{Dalem}, it is one of the important open questions of CST.

We end with a comment on the basic assumption of local finiteness. As
described in the introduction, this is motivated by the idea of a
fundamental atomicity underlying spacetime so that a finite spacetime
volume contains a finite number of causal set elements. However, local
finiteness is a stronger condition. Let us consider the anti-de Sitter
spacetime, described in \cite{HE}. This spacetime is not globally
hyperbolic, but is otherwise well behaved, satisfying stringent
causality requirements. An important global feature is that it
contains Alexandroff intervals of infinite volume. Thus, although the
number to volume correspondence can be maintained by the Poisson
sprinkling, the resulting CST-type discretisation does not produce a
locally finite poset.  While these spacetimes are perhaps not
physically realisable, they may play a non-trivial role in the
approach to classicality. Should CST relax local finiteness to a
weaker condition or must we accept that such spacetimes lose their
relevance in CST? A better understanding of causal set
geometry would be essential to help answer this question
satisfactorily.

\section{Acknowledgements} I thank the organisers of the Dagstuhl
seminar for this opportunity to explore the links between 
spacetime causality and the mathematics of posets.  In particular, I
thank Jimmy Lawson and Prakash Panangaden for discussions related to
this work. This work was supported in part by the Royal
Society-British Council International Joint Project.  
 
\end{document}